\documentclass[sigconf]{acmart}
\AtBeginDocument{%
  }

\setcopyright{acmlicensed}
\copyrightyear{2018}
\acmYear{2018}
\acmDOI{XXXXXXX.XXXXXXX}

\acmJournal{JDS}
\acmVolume{37}
\acmNumber{4}
\acmArticle{111}
\acmMonth{8}



\usepackage{enumitem}
\usepackage{multirow}
\usepackage{enumitem}

\usepackage{enumitem}
\usepackage{graphicx}
\usepackage{colortbl}
\usepackage{multirow}
\usepackage{xcolor}
\usepackage{tcolorbox} 
\usepackage{booktabs}
\usepackage{subfig}
\usepackage{lipsum}
\usepackage{algorithm}
\usepackage{algorithmic}
\usepackage{arydshln}
\usepackage{amsmath}
\usepackage{graphicx}
\usepackage{multirow}
\usepackage{adjustbox}
\usepackage{tabularx}
\usepackage{booktabs}
\usepackage{tablefootnote}
\usepackage{booktabs}
\usepackage{array}
\usepackage{balance} 
\usepackage{multirow}
\usepackage[normalem]{ulem}
\usepackage{color}
\definecolor{lightgray}{RGB}{215,215,215}
\usepackage{colortbl}  %彩色表格需要加载的宏包
\usepackage{xcolor}
\useunder{\uline}{\ul}{}
\usepackage{subcaption}
\usepackage{algorithm}  

\usepackage{amsmath}  
\usepackage{enumitem}
\usepackage{tabularx}
\usepackage[utf8]{inputenc}
\usepackage[english]{babel}
\usepackage{amsthm}
\usepackage{bm}
\usepackage{cancel}
\begin{document}

% \title{An Effective Pipeline for Pruning and Restoring LLM-Based Recommender Systems}
\title{Boosting Parameter Efficiency in LLM-Based Recommendation through Sophisticated Pruning}

\newcommand{\bkq}[1]{{\color{black}{#1}}}

\definecolor{blue-violet}{rgb}{0.54, 0.17, 0.89}
\newcommand{\zjz}[1]{{\color{black}{#1}}}

\newcommand{\todo}[1]{\textcolor{red}{\textbf{[#1]}}}

% \affiliation{%

\author{Shanle Zheng}
\authornote{Both authors contributed equally to this research.}
\affiliation{%
  \institution{University of Science and Technology of China}
  \city{Hefei}
  \country{China}}
\email{slzheng@mail.ustc.edu.cn}

\author{Keqin Bao}
\authornotemark[1]
\affiliation{%
  \institution{University of Science and Technology of China}
  \city{Hefei}
  \country{China}
}
\email{baokq@mail.ustc.edu.cn}

\author{Jizhi Zhang}
\affiliation{%
 \institution{University of Science and Technology of China}
 \city{Hefei}
 \country{China}}
\email{cdzhangjizhi@mail.ustc.edu.cn}

\author{Yang Zhang}
\affiliation{%
  \institution{National University of Singapore}
  \city{Singapore}
  \country{Singapore}}
\email{zyang1580@gmail.com}

\author{Fuli Feng}
\affiliation{%
  \institution{University of Science and Technology of China}
  \city{Hefei}
  \country{China}}
\email{fulifeng93@gmail.com}

\author{Xiangnan He}
\affiliation{%
  \institution{University of Science and Technology of China}
  \city{Hefei}
  \country{China}}
\email{xiangnanhe@gmail.com}

%% Article type: Research, Review, Discussion, Invited or position
\acmArticleType{Review}
% \acmArticleType{Research}
%%
%% Links to code and data
% \acmCodeLink{https://github.com/borisveytsman/acmart}
% \acmDataLink{htps://zenodo.org/link}
%%
%% Authors' contribution
% \acmContributions{BT and GKMT designed the study; LT, VB, and AP
%   conducted the experiments, BR, HC, CP and JS analyzed the results,
%   JPK developed analytical predictions, all authors participated in
%   writing the manuscript.}
%%
%% Sometimes the addresses are too long to fit on the page.  In this
%% case uncomment the lines below and fill them accodingly.
%%
%% \authorsaddresses{Corresponding author: Ben Trovato,
%% \href{mailto:trovato@corporation.com}{trovato@corporation.com};
%% Institute for Clarity in Documentation, P.O. Box 1212, Dublin,
%% Ohio, USA, 43017-6221}
% \keywords{Do, Not, Us, This, Code, Put, the, Correct, Terms, for,
  % Your, Paper}

% 
\begin{abstract}
    
% \end{abstract}
LLM-based recommender systems have made significant progress; however, the deployment cost associated with the large parameter volume of LLMs still hinders their real-world applications. This work explores parameter pruning to improve parameter efficiency while maintaining recommendation quality, thereby enabling easier deployment. Unlike existing approaches that focus primarily on inter-layer redundancy, we uncover intra-layer redundancy within components such as self-attention and MLP modules. Building on this analysis, we propose a more fine-grained pruning approach that integrates both intra-layer and layer-wise pruning. Specifically, we introduce a three-stage pruning strategy that progressively prunes parameters at different levels and parts of the model, moving from intra-layer to layer-wise pruning, or from width to depth. Each stage also includes a performance restoration step using distillation techniques, helping to strike a balance between performance and parameter efficiency.
Empirical results demonstrate the effectiveness of our approach: across three datasets, our models achieve an average of 88\% of the original model’s performance while pruning more than 95\% of the non-embedding parameters. This underscores the potential of our method to significantly reduce resource requirements without greatly compromising recommendation quality. Our code will be available at \url{https://github.com/zheng-sl/PruneRec}.

\end{abstract}
% \end{abstract}
\maketitle

\section{Introduction}

Recent advances in Large Language Models (LLMs)~\cite{achiam2023gpt, touvron2023llama, yang2024qwen2} have transformed recommender systems by leveraging their extensive world knowledge and advanced reasoning capabilities, establishing them as a pivotal research frontier~\cite{zhao2023survey}. 
Current research  primarily focuses on maximizing the leverage of LLMs' power to enhance recommendation performance~\cite{recsurvey1, recsurvey2, bao2023tallrec, zhang2023recommendation}, achieving remarkable progress~\cite{recsurvey4, recsurvey3}.
However, the substantial computational costs and memory overhead associated with LLMs continue to pose significant barriers to their practical deployment~\cite{chowdhery2023palm}.
  
Toward addressing this issue, we recognize that not all parameters in LLMs are necessary for recommendation computation, as LLMs are fundamentally designed for general-purpose language tasks. 
% This means, parameter redundancy exists, which can increase computation and memory requirements without offering significant utility.
That means there is parameter redundancy, which could lead to a redundant burden of computation and memory without offering significant utility.
In this light, we aim to improve parameter efficiency to mitigate the issue of computational and memory overhead.
The goal is twofold: to reduce parameter volume and to preserve recommendation quality, ultimately enabling space- and time-efficient recommendations that maintain high quality for real-world deployment.
% bridging the gap between LLM-based recommender systems and the deployment requirements of recommender systems.

Prior research in recommendation has primarily focused on two paradigms to improve parameter efficiency. The first paradigm uses knowledge distillation to transfer the capabilities of LLMs to lightweight models with a significantly smaller parameter scale~\cite{cui2024distillation}. While effective for parameter reduction, this approach often struggles to fully capture the depth of LLM knowledge due to the capacity gap between the two models, resulting in suboptimal performance. The second line of research applies model pruning techniques to directly reduce the LLMs' parameters, thus avoiding the introduction of a capacity gap. However, existing works typically rely on layer-wise pruning~\cite{xu2024slmrec}, which is overly coarse and overlooks the potential redundancies within each remaining layer.

This work explores the feasibility of introducing parameter pruning at a more fine-grained level.
We conduct a series of preliminary experiments, confirming that LLM-based recommender systems not only exhibit layer-wise redundancy but also redundancy within layers, as evidenced by a small number of high-magnitude activations in certain self-attention and MLP layers significantly impacting the model output, while others have minimal effects.
Based on this, we propose a new pruning pipeline that emphasizes fine-grained importance estimation and pruning of intra-layer parameters, going beyond merely layer-wise pruning, namely \textit{\textbf{PruneRec}}.
Considering the complex relationships between different levels and parts of parameters, pruning parameters across multiple levels and parts in a single step could directly lead to model collapse. 
Instead, our PruneRec follows an iterative approach: in each iteration, redundant parameters at a specific level (either layer-wise or intra-layer) or within certain components (attention or MLPs) are pruned, followed by a performance restoration step to recover the model’s capacity and prepare it for the next round of parameter importance estimation via distillation techniques.

Specifically, our pipeline consists of three stages. In the first stage, we reduce the model width by pruning attention heads and vocabulary embeddings. For attention heads, we propose a dynamic weighting algorithm that integrates influence from shallow to deep layers, combined with a KL divergence loss, to effectively quantify the importance of each head and accordingly prune the unimportant ones. 
For embedding dimensions, we prune embeddings corresponding to less important vocabulary tokens, where importance is determined based on token relevance and gradient magnitude.
In the second stage, we address redundancy in MLPs by retaining only the top 10\% of rows or columns based on importance scores defined according to the activation strengths.
Finally, in the third stage, we perform layer-level pruning to address inter-layer redundancy, retaining only the most important layers, where each layer's importance is measured by evaluating the performance difference between with and without the layer.
To ensure accurate importance estimation for the subsequent stage and maintain final performance, we also apply knowledge distillation at the end of each stage, using reverse KL divergence and cross-entropy loss to restore the pruned model's capability.

We conduct experiments on three publicly available datasets and demonstrate that our method outperforms conventional models, as well as pruning-based and distillation-based LLM approaches in recommendation, across all evaluated benchmarks. The proposed framework retains 88\% of the original LLM's performance while utilizing only 5\% of its parameters (excluding embeddings), significantly improving memory and inference efficiency, and thereby reducing deployment complexity.

To sum up, our key contributions are:
\begin{itemize}

\item We highlight that existing LLM-based recommender systems exhibit both significant intra-layer and layer-wise parameter redundancy, severely limiting the model efficiency.

% \item We highlight that existing LLM-based recommender systems exhibit significant parameter redundancy in their structures, which severely impacts the efficiency of the models.

\item We propose \textbf{PruneRec}, an iterative prune-and-restore pipeline that incorporates both layer-wise and intra-layer pruning to achieve more effective pruning while preserving recommendation performance.

% \item We propose \textit{\textbf{PruneRec}}, a prune-and-restore pipeline specifically designed for LLM-based recommender systems, which systematically optimizes the trade-off between model parameter efficiency and performance preservation.

\item Extensive experiments across multiple benchmarks demonstrate that our framework maintains competitive recommendation performance while retaining less than 5\% of the original parameters (excluding embeddings), empirically validating its effectiveness.

\end{itemize}

\section{Preliminary}

In this section, we will first briefly introduce the definitions of LLM-based sequential recommendation (\S\ref{sec:2.1}). Then, we will demonstrate, through observational experiments (\S\ref{sec:2.2}), that LLM-based recommendation models contain a significant amount of parameter redundancy, and we are eager to address this issue through model pruning (\S\ref{sec:2.3}).
\subsection{LLM-based  sequential Recommendation}
\label{sec:2.1}

Currently, using LLMs for sequential recommendation is a hot topic. Researchers aim to improve item understanding by integrating world knowledge from LLMs, enhancing recommendation performance. For a user with interaction history $\mathcal{I} = [i_1, i_2, \ldots, i_n]$, natural language prompts are constructed as input to the LLM, and the model is tasked to maximize the probability of generating the textual representation $\mathcal{Y}_{i_{n+1}}$ of the target item $i_{n+1}$. This is formalized as:
\begin{equation}
    \arg\max_{\mathcal{M}} P_{\mathcal{M}}(\mathcal{Y}_{i_{n+1}} \mid \text{Prompt}(\mathcal{I})),
\end{equation}

where $\text{Prompt}(\mathcal{I})$ converts the instruction and interaction history into natural language.

However, since LLMs are not designed for recommendations, fine-tuning is often required to align them with task-specific needs. This involves adjusting model parameters by minimizing Cross-Entropy Loss, ensuring outputs better suit the recommendation task:
\begin{equation}
\mathcal{L}_{\text{CE}} = -\sum_{i_{n+1}} \log P_{\mathcal{M}}(\mathcal{Y}_{i_{n+1}} \mid \text{Prompt}(\mathcal{I})).
\end{equation}

While fine-tuning improves performance, it often diminishes general-purpose capabilities encoded in a significant portion of the model's parameters, increasing the redundancy of the model.  We will explore this issue further and propose solutions in subsequent sections.

\subsection{Observation Experiments}
\label{sec:2.2}

In this part, we will discuss our observation experiments on the Amazon Video Games datasets. In summary, it is guided by three questions:  
\textbf{Q1}: Does a LLM fine-tuned on recommendation domain data exhibit significant parameter redundancy?  
\textbf{Q2}: From a structural perspective, in which parts of the model do redundant parameters exist?  
\textbf{Q3}: Can techniques from general LLMs be directly applied to prune redundant parameters in recommendation models effectively?

\begin{figure}[t]
  \centering
  \includegraphics[width=0.45\textwidth]{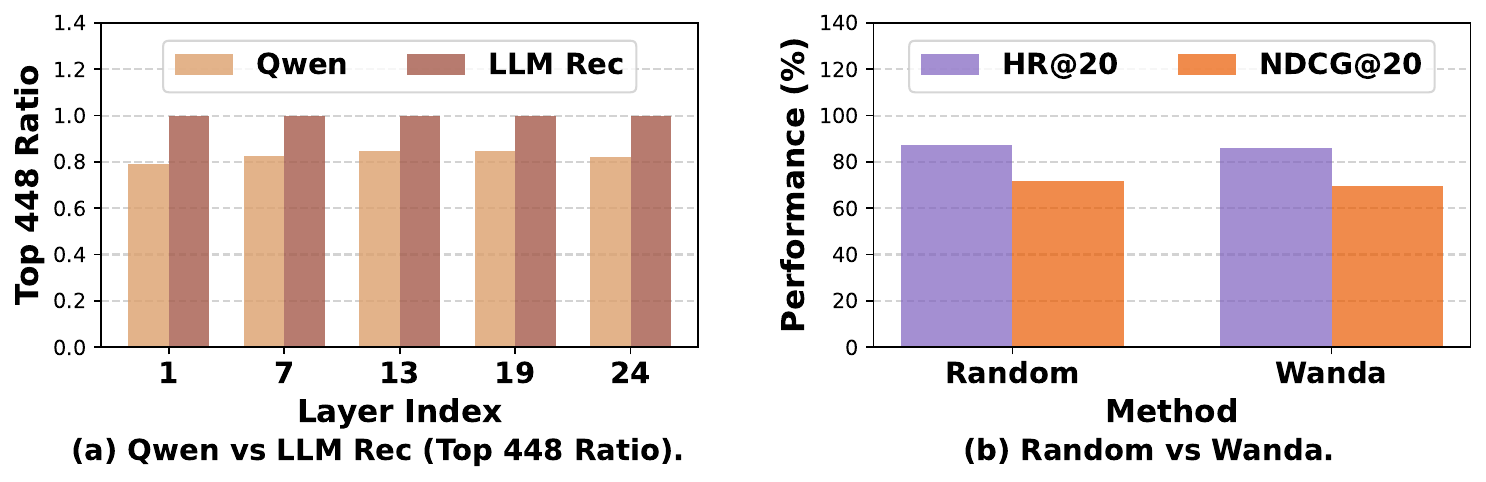} % 图片路径
  \caption{Figure (a) shows the ratio of head activations to total activations in the MLP hidden states for both the LLM (Qwen) and the LLM-based recommender system. A higher ratio indicates a greater likelihood of redundancy. Figure (b) compares the performance retention rates of random pruning and the WANDA~\cite{sun2023simple} when applied to pruning in LLM-based recommender systems.}

  \label{fig:qwen_vs_rec}                                       
\end{figure}

\paragraph{\textbf{Q1:}} 

To address this question, we randomly sampled 100 samples from the dataset and fed them into the LLM. 
We then computed the proportion of the top K\% largest activation values (i.e., the ratio of the sum of their absolute values to the total sum of absolute values across all dimensions) for each layer. This metric quantifies the distributional differences across layers. Specifically, this ratio reflects the concentration of information within each layer: a higher ratio indicates that the information is predominantly concentrated in a small number of dimensions, suggesting greater redundancy in those dimensions within the model.

The results, shown in Figure~\ref{fig:qwen_vs_rec}, compare the changes in metrics before and after fine-tuning. Compared to the fine-tuned model, the proportion of top activation values is significantly higher. This suggests that during the fine-tuning process, the LLM tends to strengthen certain capabilities that are beneficial for the recommendation task. Consequently, this leads to a structural divergence in the internal weight distribution, with certain dimensions gradually dominating. This finding supports our assumption: after fine-tuning on recommendation-specific data, the model primarily relies on parameters specialized for the recommendation task, while parameters supporting general-purpose capabilities are weakened. As a result, the fine-tuned model exhibits greater parameter redundancy compared to the general-purpose model.

\paragraph{\textbf{Q2:}} 
Previous studies have demonstrated that LLMs exhibit significant redundancy across layers in the recommendation scenario~\cite{xu2024slmrec}. To further investigate structural redundancies in LLMs, we focus on the two most critical components of the Transformer architecture: the self-attention mechanism and the multi-layer perceptron layer. We conduct a detailed analysis of these two modules in an LLM fine-tuned on recommendation data. Specifically, we randomly sampled 100 instances from the dataset and fed them into the LLM fine-tuned for recommendation tasks. We then examined the proportion of activation values contributed by the top K\% of dimensions in the output of the attention layer and the activations of the intermediate dimensions in the MLP layer. 

\begin{figure}[t]
  \centering
  \includegraphics[width=0.5\textwidth]{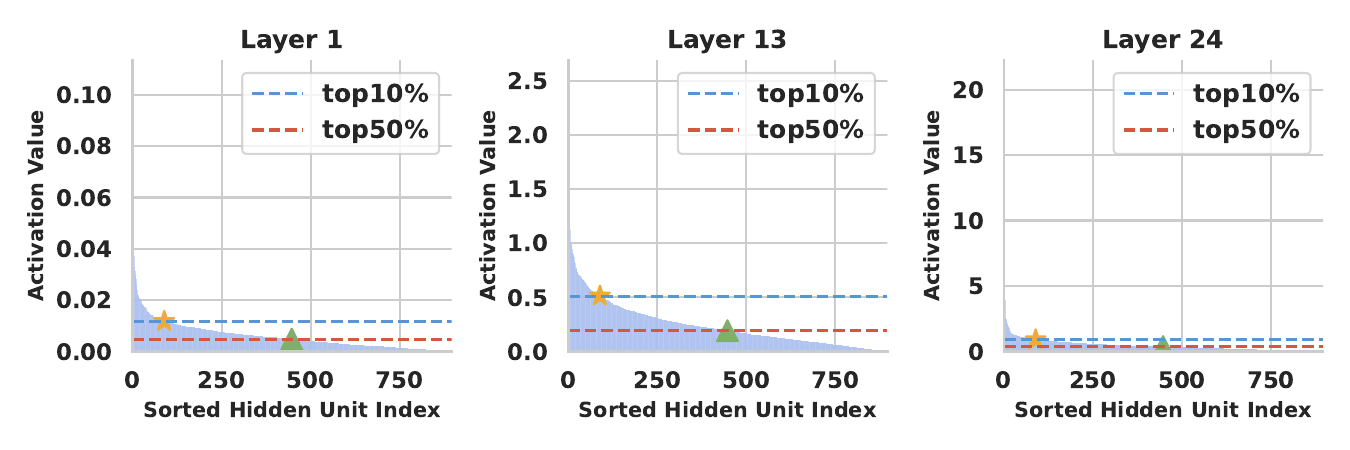} % 图片路径
  \caption{This figure shows the distribution of activation magnitudes in the attention layers of the 1st, 13th, and 24th layers of the model. We have marked the values corresponding to the top 10\% and top 50\% activations in the figure.}                % 图片标题
  \label{fig:attn_output}                                        % 标签（必须放在caption后）
\end{figure}

\begin{figure}[t]
  \centering
  \includegraphics[width=0.5\textwidth]{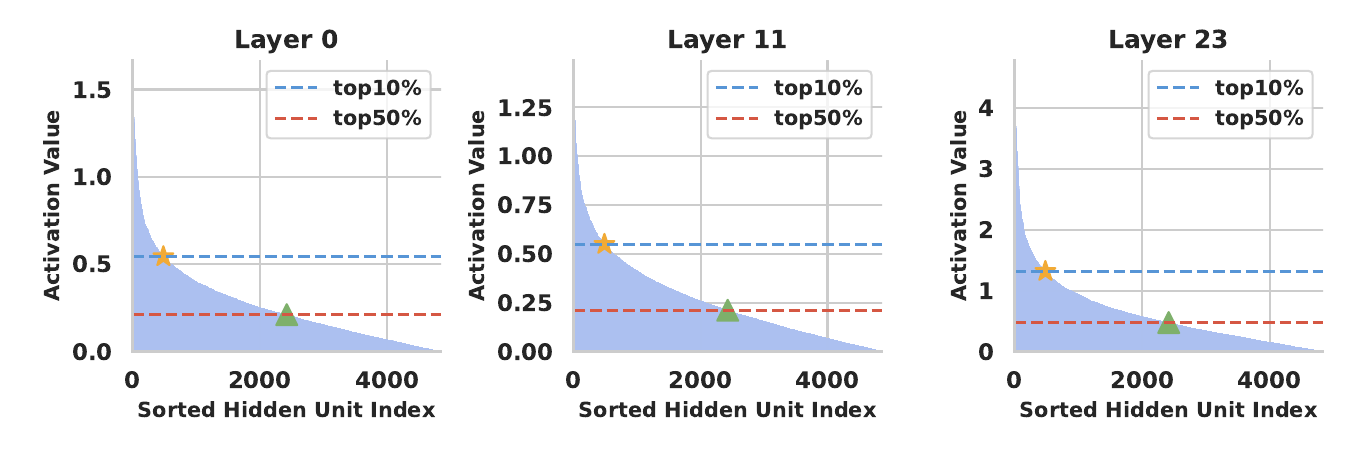} % 图片路径
  \caption{This figure illustrates the distribution of activation magnitudes in the intermediate layers of the MLP for the 1st, 13th, and 24th layers of the model. We have marked the values corresponding to the top 10\% and top 50\% activations in the figure.}                % 图片标题
  \label{fig:mlp_output}                                        % 标签（必须放在caption后）
\end{figure}

Consistent with prior findings, as shown in the Figure~\ref{fig:attn_output} and Figure~\ref{fig:mlp_output}, the activations of the model exhibit a significant long-tail effect in both the MLP and Attention layers. A higher proportion of activation in specific dimensions indicates that these dimensions play a more dominant role, which also suggests greater redundancy within the corresponding module. This finding highlights significant parameter redundancy in these two components of the LLM-based recommendation system~\cite{zhu2024collaborative,wang2024rethinking}. Given this observation, there is an urgent need to explore strategies for identifying and removing redundant parameters to improve model efficiency and performance.

\paragraph{\textbf{Q3:}} To address this issue, an intuitive approach is to directly adapt model pruning methods commonly used in the general LLM domain. Following this idea, we applied WANDA~\cite{sun2023simple}, a widely-used pruning technique, which evaluates the importance of each weight by computing the Hadamard product of the absolute value of the weight and the norm of its corresponding input activation. This method effectively prunes less important parameters, reducing the model size and computational cost while minimizing performance degradation. However, as shown in Figure~\ref{fig:qwen_vs_rec}, our observation experimental results reveal that, despite its strong performance on general tasks, WANDA underperforms random pruning when applied to LLM-based recommender systems. This suggests that pruning techniques optimized for general tasks do not necessarily preserve performance effectively in recommendation scenarios. Therefore, to better retain recommendation capabilities, it is essential to design a specialized pruning framework tailored for recommendation models.

\subsection{Model Pruning}
\label{sec:2.3}

Before delving into the specifics of pruning methods, we first provide a brief introduction to model pruning. In detail, in the domain of general LLMs, this technique identifies and removes weights or neurons that contribute minimally to the model's output. By pruning the model, the total number of parameters can be significantly reduced, leading to lower storage requirements, faster inference, and improved deployment capabilities in resource-constrained environments.
Its primary approaches include unstructured pruning~\cite{sun2023simple}, structured pruning~\cite{wang2019structured, ma2023llm, an2024fluctuation, zhong2024blockpruner}, and dynamic parameter pruning~\cite{anagnostidis2023dynamic}. Unstructured pruning, while effective in reducing redundancy, results in irregular weight distributions that are difficult to deploy efficiently on existing hardware. Dynamic pruning, on the other hand, adaptively selects subsets of active parameters during inference. However, in the context of recommendation systems, certain parameters supporting tasks like math or code reasoning are not essential and can be entirely removed. Therefore, in this work, we focus on structured pruning, which removes entire neurons~\cite{ma2023llm}, channels~\cite{hou2025instruction}, or layers~\cite{men2024shortgpt}. This approach generates more regularized models that can be directly accelerated using existing deep learning frameworks and hardware accelerators, making it particularly suitable for real-world deployment in recommendation scenarios.

% \begin{figure}[htbp]
%   \centering
%   \includegraphics[width=0.5\textwidth]{samples/picture/attn_o_layer_23_activation.png} % 图片路径
%   \caption{self-attention layer's output in LLM-based RS model}                % 图片标题
%   \label{fig:attn_output}                                        % 标签（必须放在caption后）
% \end{figure}

\section{PruneRec Pipeline}
This section provides a detailed overview of our proposed method, which consists of two key components: pruning and restoration. For the pruning, motivated by our preliminary experimental findings, we remove redundant parameters from the LLM-based recommendation model that are irrelevant to its recommendation capabilities. For the restoration, we recover the model's performance through knowledge distillation.

Specifically, we first follow BIGRec~\cite{bao2023bi} to train on the corresponding recommendation dataset, followed by pruning and distillation. Our pipeline consists of three stages, during each of which we prune a portion of the model's parameters and utilize the original model to assist in recovering performance. In the following sections, we will detail how our three stages perform pruning (\S\ref{sec:3.1}, \S\ref{sec:3.2}, \S\ref{sec:3.3}), and then introduce our method for restoring performance (\S\ref{sec:3.4}).

\subsection{Stage I}
% 在第一阶段我们首先尝试裁剪模型的宽度，具体来说我们首先对self-attention模块和model dimension进行裁剪，旨在减少模型的维度同时减少attention中的head数。
In the first stage, we initially attempt to prune the model's width. Specifically, we start by pruning the self-attention module and the model dimension, aiming to reduce the model's dimensionality while also decreasing the number of heads in the attention mechanism.
\subsubsection{self-attention module}
% \subsubsection{111}
% \subsubsubsection{222}

\label{sec:3.1}
For the self-attention module, we begin by assessing the importance of each attention head. Given an input matrix $X \in \mathbb{R}^{S \times d_{\text{model}}}$, where $ S $ is the sequence length and $ d_{model} $ is the model's hidden dimension, the attention module computes queries ($Q$), keys ($K$), and values ($V$) for $h$ heads using independent linear transformations. For the $i$-th head, the attention score matrix $A_i$ is computed as:
\[
A_i = \text{Softmax}\left(\frac{Q_i K_i^T}{\sqrt{d_k}}\right),
\]
where $d_k = d_{\text{model}} / h$ ensures numerical stability. The output of each head is then obtained as $O_i = A_i V_i$. This process forms the basis for evaluating the contribution of individual attention heads to the model's performance.

Then, following prior work~\cite{zhou2024role,deng2025cram}, we define the perturbed output of the $i$-th head in layer $l$ as:
\begin{equation}
\hat{O}_i^l = \text{Softmax}\left(\frac{\epsilon_i^l Q_i K_i^T}{\sqrt{d_k}}\right) V_i,
\end{equation}
where $\epsilon_i^l$ is a near-zero scalar introduced to suppress the ability of the $i$-th head in layer $l$ to capture input information.

After that, we propagate the perturbed output $\hat{O}_i^l$ through the remaining layers to obtain the corresponding output logits. If an attention head is unimportant, the output logits when $\epsilon_i^l \to 0$ should closely resemble the original logits. To quantify this effect, we measure the importance of each attention head by computing the Kullback–Leibler divergence between the output distributions before and after setting $\epsilon_i^l \to 0$. Specifically, the importance of attention head $O_i^l$ is defined as:

\begin{equation}
\text{Imp}(O_i^l) = D_{\text{KL}}\big(p(x, O_i^l) \,\|\, p(x, \cancel{\hat{O}_i^l})\big),
\end{equation}
where $p(x, O_i^l)$ represents the output distribution with the $i$-th head intact, and $p(x, \cancel{\hat{O}_i^l})$ represents the output distribution when the $i$-th head is suppressed.

To ensure comparability across heads, we normalize the raw importance scores within each layer using min-max normalization:
\begin{equation}
\text{Imp}(\text{O}_i^l) = \frac{\text{Imp}(\text{O}_i^l) - \min\big(\text{Imp}(\text{O}^l)\big)}{\max\big(\text{Imp}(\text{O}^l)\big) - \min\big(\text{Imp}(\text{O}^l)\big)}.
\end{equation}

Furthermore, we recognize that the output of earlier layers in a Transformer influences the output of subsequent layers. By analogy, the importance of attention heads in earlier layers propagates to affect the importance of heads in later layers. To capture this dependency, we propose a recursive importance computation strategy. For the $i$-th head in layer $l+1$, its final importance score is determined by:
\begin{equation}
\text{Imp}(\text{O}_i^{l+1}) = \alpha \cdot \text{Imp}(\text{O}_i^l) + (1 - \alpha) \cdot \text{Imp}(\text{O}_i^{l+1}),
\end{equation}
where $\alpha \in [0, 1]$ controls the propagation weight from preceding layers.

In practice, we compute importance scores layer-by-layer using an iterative procedure. This approach avoids the combinatorial explosion of $O(2^{L \times h})$ possibilities (for $L$ layers and $h$ heads per layer), ensuring computational tractability. Finally, we randomly select $B$ samples from the training set to compute the importance scores of the attention heads and perform pruning based on these scores to remove $K_{Attn}$ heads for each layer~\footnote{In practice, we set $K_{Attn}$ = 7 in our main experiments}.

\subsubsection{Embedding Layer}
\label{sec:3.3}
To mitigate the influence of low-relevance tokens (e.g., stopwords), we evaluate importance by combining the magnitude of the embedding weights with their gradients. 
Specifically, we randomly the sample $B$ samples and compute the gradient of the embedding weights via backpropagation, denoted as \( \nabla_E \). The element-wise product \( E \odot \nabla_E \) is then calculated, and its absolute value is used as the importance score for each embedding weight. In practice, we use the following formula:
\begin{equation}
    I_d = \frac{1}{B \cdot S} \sum_{b=1}^{B} \sum_{s=1}^{S} \left| E_{d} \odot \nabla_{E_d} \right|_{(b,s)},
\end{equation}
where \( \left| E_{d} \odot \nabla_{E_d} \right|_{(b,s)} \) represents the absolute value of the element-wise product of the embedding weights and their gradients for the \( d \)-th dimension at sample \( b \) and sequence position \( s \).
Based on the computed importance scores \( \{I_1, I_2, \cdots, I_{d_{model}}\} \), we rank the dimensions and retain those matching the input dimension of the attention layer. The embedding matrix is then pruned by removing rows corresponding to the unselected dimensions.

\subsection{Stage II}
\label{sec:3.2}
% In prior experiments, we observed that the activations of intermediate dimensions in MLP layers exhibit significant sparsity. This suggests that many dimensions contribute minimally to feature representation. Motivated by this observation, we propose a magnitude-based pruning method to remove redundant weights from the MLP layer in the stage II. Specifically, we focus on the actual contribution of each activation dimension to downstream recommendation tasks. A parameter can be safely pruned only if its corresponding activation dimension has negligible impact on the final prediction.
In prior experiments, we observed significant sparsity in the activations of intermediate dimensions within MLP layers, indicating that many dimensions contribute minimally to feature representation. Based on this insight, we propose a magnitude-based pruning method to remove redundant weights from the MLP layer in Stage II. Specifically, we evaluate the actual contribution of each activation dimension to downstream recommendation tasks, ensuring that only parameters with negligible impact on the final prediction are pruned.

Given an input sequence $ X \in \mathbb{R}^{S \times d_{model}} $, the MLP layer processes features through a three-stage transformation and can be expressed as:
\begin{equation}
    Y = \left( \sigma(XW_{\text{gate}} + b_{\text{gate}}) \odot (XW_{\text{up}} + b_{\text{up}}) \right) W_{\text{down}} + b_{\text{down}},
\end{equation}
where $ \odot $ denotes element-wise multiplication, $ W_{\text{gate}}, W_{\text{up}}, W_{\text{down}} $ are weight matrices, and $ b_{\text{gate}}, b_{\text{up}}, b_{\text{down}} $ are bias terms.

To prune the redundant parameters, we introduce sparsity constraints to quantify the importance of each intermediate dimension for recommendation tasks and prune redundant dimensions.  Given the up-projected output $H = XW_{\text{up}} + b_{\text{up}}$, we extract the activation values corresponding to the last token of each sequence:
\begin{equation}
    H_{\text{last}} = H[-1, :] \in \mathbb{R}^{d_{\text{up}}}.
\end{equation}
Then, for each dimension $ d \in \{1, 2, \ldots, d_{\text{up}}\} $, we compute the frequency with which the activation magnitude exceeds a predefined threshold $\tau$ across $B$ randomly selected training samples:

\begin{equation}
    S(d) = \sum_{b=1}^{B} \mathbb{I}\left( \left| H_{\text{last}}^{(b,d)} \right| > \tau \right),
\end{equation}
where $ \mathbb{I}(\cdot) $ is an indicator function.We then retain the top-$K$ most frequently activated dimensions by computing:
\begin{equation}
I = \underset{d \in \{1, \ldots, d_{\text{up}}\}}{\text{argtopK}}(S(d), K_{MLP}),
\end{equation}
where $ \text{argtopK}(\cdot, K_{MLP}) $ returns the indices of the $ K_{MLP} $ dimensions with the highest activation frequency\footnote{In practice, we set $K_{MLP}$=896 for a moderate performance}.

Based on the selected dimension set $ I $, we finally apply structured pruning to the corresponding rows or columns of the weight matrices $ W_{\text{up}}, W_{\text{gate}}, $ and $ W_{\text{down}} $. This method allows for compressing the intermediate dimension to a target sparsity rate while keeping the input and output dimensions unchanged.

\subsection{Stage III}
\label{sec:3.4}

In the previous two stages, we discussed how to prune intra-layer modules to reduce parameter redundancy. However, excessive pruning of a single layer may inadvertently remove critical parameters, harming model performance. In this section, we propose a method for layer-level pruning to address inter-layer redundancy.Following prior work~\cite{zhong2024blockpruner}, we first evaluate the change in perplexity (PPL) on a calibration set when each layer is removed. A larger PPL increase indicates higher layer importance. To quantify this, we use the following formula:
\begin{equation}
    \Delta \text{PPL}_l = \text{PPL}_{\text{masked}(l)} - \text{PPL}_{\text{original}},
\end{equation}
where \( \text{PPL}_{\text{masked}(l)} \) is the perplexity after masking layer \( l \), and \( \text{PPL}_{\text{original}} \) is the original model's perplexity. This score measures each layer's contribution: higher scores indicate greater importance.
After computing importance scores for all layers, we iteratively remove the least important layer until the model reaches $K_{layer}$ layers\footnote{In practice, we set $K_{Layer}$=16.}. This approach effectively reduces inter-layer redundancy while preserving critical layers for model performance.
\subsection{Restoration}

After completing the pruning process in each stage, we aim to recover the model's performance.

This offers two benefits: firstly, it allows for verifying the effectiveness of pruning; secondly, since pruning different modules may involve conflicting interactions, restoring performance helps more accurately measure the importance metrics for subsequent module pruning.
To achieve this, we leverage the original model's output logits and use the training set for knowledge distillation to refine the pruned model. Specifically, we employ the pre-pruned model as the teacher model and minimize the forward Kullback-Leibler (KL) divergence between the outputs of the pruned and original models. Additionally, to address potential biases in the teacher model's logits, we incorporate a cross-entropy (CE) loss with respect to the ground-truth labels for better balance. The total loss function is formulated as follows:
\begin{equation}
    L_{\text{total}} = \lambda \cdot \text{KL}(p_T(x) \| p_S(x)) + (1-\lambda) \cdot \text{CE}(y, p_S(x)),    
\end{equation}

where $ p_T(x) $ and $ p_S(x) $ represent the probability distributions obtained by applying softmax to the logits of the teacher and student models, respectively, $ y $ denotes the ground-truth label, and $ \lambda \in [0, 1] $ is a balancing hyperparameter.

\begin{table}[t]
\centering
\begin{tabular}{l|ccc}
\toprule
      & Games & CDs   & Sports \\ \midrule
Item  & 11037 & 14239 & 16003  \\
Train & 201613 & 148685 & 181477 \\
Valid & 25202 & 18586 & 22685  \\
Test  & 25203 & 18587 & 22686  \\
\bottomrule
\end{tabular}
\caption{The table presents statistical information for three datasets. The 1st row shows the number of items, and the 2nd, 3rd, and 4th rows display the number of sequences in the training, validation, and test sets, respectively.}
\label{table:statstic}
\vspace{-2em}
\end{table}

\begin{table*}[htbp]
  \centering
  \small
  \renewcommand{\arraystretch}{1.1}
  \caption{Recommendation accuracy of the compared methods on cross-domain datasets. ``\textit{Traditional Recommender Systems}'' refers to conventional methods that rely on ID-based approaches, while ``\textit{Recommender Systems Utilizing LLMs}'' refers to methods that leverage LLMs and specifically aim to improve parameter efficiency. \textbf{Param} indicates the non-embedding parameter count of the model, and \textbf{Avg} calculates the average performance of the method across 12 metrics on 3 datasets, expressed as a percentage relative to LLM-based methods without parameter pruning.}
  \label{tab:main_results}
  \vspace{-0.5em}
  \scalebox{0.77}{
\begin{tabular}{llcccccccccccccc}
    \toprule
\multicolumn{2}{l}{Model} & Param                                                         & \multicolumn{4}{c}{\textbf{Games}}                                    & \multicolumn{4}{c}{\textbf{Sports}}                                   & \multicolumn{4}{c}{\textbf{CDs}}                                      & \textbf{Avg(\%)} \\ \cline{1-16}
\multicolumn{3}{l}{}                                                                & \textbf{HR@10} & \textbf{NDCG@10} & \textbf{HR@20} & \textbf{NDCG@20} & \textbf{HR@10} & \textbf{NDCG@10} & \textbf{HR@20} & \textbf{NDCG@20} & \textbf{HR@10} & \textbf{NDCG@10} & \textbf{HR@20} & \textbf{NDCG@20} & \textbf{}        \\ \hline
\multicolumn{2}{l}{BIGRec } & 357M                                               & 0.0852         & 0.0520           & 0.1200         & 0.0608           & 0.1199         & 0.0993           & 0.1407         & 0.1046           & 0.1094         & 0.0771           & 0.1299         & 0.0823           & 100.0            \\ 
\cdashline{1-16}
\multicolumn{15}{c}{\textit{Traditinoal Recommender Systems}} \\
\multicolumn{2}{l}{GRU4Rec } & 0.7M                                                 & 0.0374         & 0.0205           & 0.0603         & 0.0262           & 0.0785         & 0.0629           & 0.0908         & 0.0660           & 0.0492         & 0.0308           & 0.0655         & 0.0349           & 52.7             \\
\multicolumn{2}{l}{Caser }  & 1.8M                                                  & 0.0316         & 0.0166           & 0.0512         & 0.0215           & 0.0717         & 0.0523           & 0.0862         & 0.0559           & 0.0416         & 0.0246           & 0.0593         & 0.0291           & 45.9             \\
\multicolumn{2}{l}{SASRec}   & 4.5M                                               & 0.0578         & 0.0358           & 0.0818         & 0.0418           & 0.0875         & 0.0753           & 0.0979         & 0.0779           & 0.0843         & 0.0598           & 0.1005         & 0.0638           & 73.2             \\
\cdashline{1-16}
\multicolumn{15}{c}{\textit{Recommender Systems Ultilizing LLMs}} \\
\multicolumn{2}{l}{DLLM2Rec }   & 4.5M                                          & 0.0721         & 0.0421           & 0.1056         & 0.0505           & 0.0994         & 0.0776           & 0.1170         & 0.0821           & 0.0802         & 0.0567           & 0.1000         & 0.0616           & 80.0             \\
\multicolumn{2}{l}{SLMRec}   & 14.9M                                              & 0.0430         & 0.0238           & 0.0721         & 0.0311           & 0.0890         & 0.0730           & 0.1045         & 0.0769           & 0.0737         & 0.0473           & 0.0931         & 0.0522           & 63.8             \\ 
\cdashline{1-16}
% \multirow{3}{*}{Ours} & \rowcolor[HTML]{EFEFEF} Stage1 (178M)  & 0.0795         & 0.0468           & 0.1166         & 0.0562           & 0.1094         & 0.0902           & 0.1324         & 0.0960           & 0.1000         & 0.0698           & 0.1204         & 0.0749           & 92.4             \\
%                       & \rowcolor[HTML]{E4E4E4} Stage2 (51.0M) & 0.0827         & 0.0487           & 0.1191         & 0.0579           & 0.1078         & 0.0894           & 0.1305         & 0.0951           & 0.1000         & 0.0694           & 0.1185         & 0.0741           & 92.5             \\
                      \rowcolor[HTML]{EFEFEF} \multicolumn{2}{l}{PruneRec (ours) } & 17.0M & \textbf{0.0803}         & \textbf{0.0475}           & \textbf{0.1164}         & \textbf{0.0566}           & \textbf{0.1009}         & \textbf{0.0827}           & \textbf{0.1198}         & \textbf{0.0875}           & \textbf{0.0946}         & \textbf{0.0658}           & \textbf{0.1156}         & \textbf{0.0712}           & \textbf{88.6}             \\ \hline
\end{tabular}}
\vspace{-1em}
\end{table*}

\begin{figure*}[htbp]
  \centering
  \includegraphics[width=0.9\textwidth]{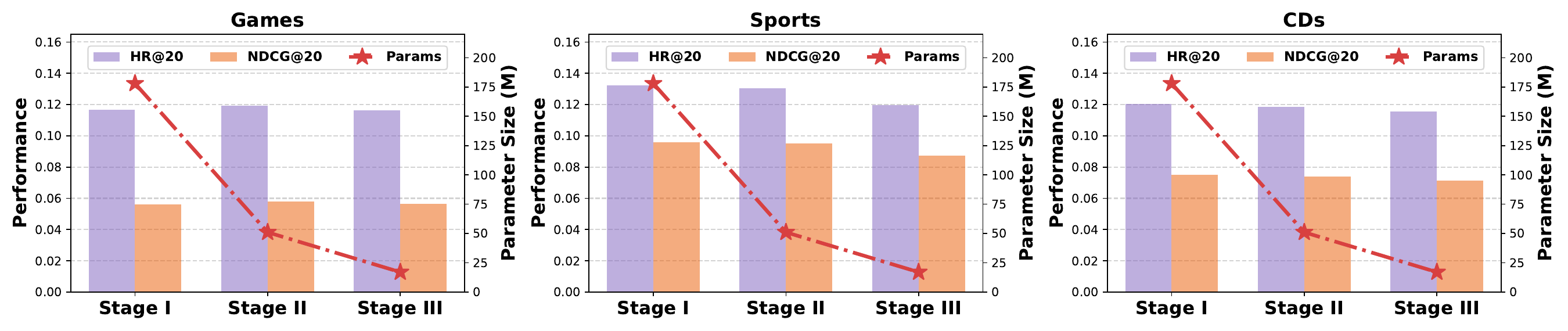} % 图片路径
  \vspace{-1em}
  \caption{This figure illustrates the changes in performance and parameter count across the three stages of our method for different datasets. In each figure, the left vertical axis quantifies the metrics NDCG@20 and HR@20, while the right vertical axis measures the parameter size. The bar charts represent the NDCG@20 and HR@20 metrics, and the dashed line depicts the trend in parameter count reduction.}                % 图片标题
  % 这张图展示了不同数据集在我们方法的三个阶段的性能和参数量的变化，每张图左边的纵轴量化了NDCG@20和HR@20的指标，右边的纵皱量化了参数量的大小，柱状图对应NDCG@20和HR@20的指标，折线部分对应参数量的变化趋势
  \label{fig:rq1}                                        % 标签（必须放在caption后）
\end{figure*}

\section{Experiments}
In this section, we compare the PruneRec approach with baseline methods, including traditional recommendation models and LLM-based systems. We evaluate its effectiveness on three Amazon datasets—Video Games, Sports, and CDs—through comparative experiments and analyze the contribution of individual components in our methodology.
\subsection{Experimental Settings}  
\subsubsection{Datasets}  
We perform experiments on three real-world Amazon review datasets: Video Games, Sports, and CDs. Following prior work~\cite{bao2024decoding}, we preprocess the data by retaining only users with at least five interactions, organizing their historical records chronologically, and capping each user's history to a maximum of 10 entries. To simulate real-world temporal dependencies, we partition the datasets into training, validation, and test sets in an 8:1:1 ratio based on timestamps. This chronological split ensures temporal consistency and prevents data leakage~\cite{ji2023critical}. Dataset statistics are summarized in Table~\ref{table:statstic}.

\subsubsection{Evaluation}
To evaluate recommendation performance, we use two key metrics: HR@K and NDCG@K. We set K in [10, 20]. For a more comprehensive analysis of each component's contribution, we conduct detailed ablation experiments.
\subsubsection{Baselines}
% \begin{itemize}[leftmargin=*, nosep]
- \textbf{SASRec~\cite{kang2018self}}  leverages attention mechanisms to model users' historical interactions and recommends the next item that a user may be interested in based on this information.
- \textbf{GRU4Rec~\cite{hidasi2015session}} is an RNN-based approach that utilizes GRUs to capture users' historical interactions and generate recommendations accordingly.
- \textbf{Caser~\cite{tang2018personalized}}  comprehensively models user behaviors using both horizontal and vertical convolutional operations to capture diverse sequential patterns.
- \textbf{DLLM2Rec~\cite{cui2024distillation}} employs knowledge distillation to transfer recommendation capabilities from LLMs to traditional recommendation systems, bridging the gap between advanced and conventional models.
- \textbf{SLMRec~\cite{xu2024slmrec}} identifies layer redundancy in LLM-based recommendation models and applies a distillation strategy to transfer knowledge from a deeper teacher model to a shallower student model, ensuring efficiency without compromising performance.
% \end{itemize}

\subsubsection{Implementing Details}%已修改
For traditional recommendation models, we optimize them using the binary cross-entropy loss~\cite{ruby2020binary} and the Adam optimizer with a learning rate selected from $\{10^{-2}, 10^{-3}, \\ 10^{-4}\}$. The data is processed in batches of size $1024$, while the weight decay is tuned within the range $\{10^{-2}, 10^{-3}, 10^{-4}, 10^{-5}, 10^{-6} \\\}$. Additionally, the embedding size is adjusted among $\{64, 128, 256\}$.
For SLMRec, we trim the number of layers to be comparable to our parameter count for a fair comparison. Additionally, we adjust the loss function during distillation, as our experiments demonstrate that our proposed loss yields better performance.
For our method, we use Qwen2-0.5B~\cite{yang2024qwen2} as the base model. First, we follow BIGRec~\cite{bao2023bi}, fine-tune the LLM on the datasets for 3 epochs with a batch size of 128 and a learning rate of $3 \times 10^{-4}$. We then adjust the hyperparameter $\alpha$ within the range $[0, 1]$ in intervals of $0.1$. To balance the KL loss and CE loss during distillation, we adjust $\lambda$ in $[0, 1]$ in intervals of $0.2$.

% We use Qwen2-0.5B as the base model for our experiments. We first fine-tune the LLM model on the pre-processed datasets to ensure that the LLM exhibits strong recommendation capabilities. Each model is trained for 3 epochs with a batch size of 128, and the learning rate is set to 3e-4. Regarding the hyperparameters used in the method section, we set different $\alpha$ values for different datasets to determine the influence of shallow layers on deeper layers. Specifically, for the Games dataset, $\alpha$ is set to 0.7, for the Sports dataset, $\alpha$ is set to 0.2, and for the CDS dataset, $\alpha$ is set to 1.0. To balance the KL loss and the cross-entropy (CE) loss during the distillation process, we set the $\lambda$ parameter to 0.8.

\begin{figure*}[htbp]
  \centering
  \includegraphics[width=0.9\textwidth]{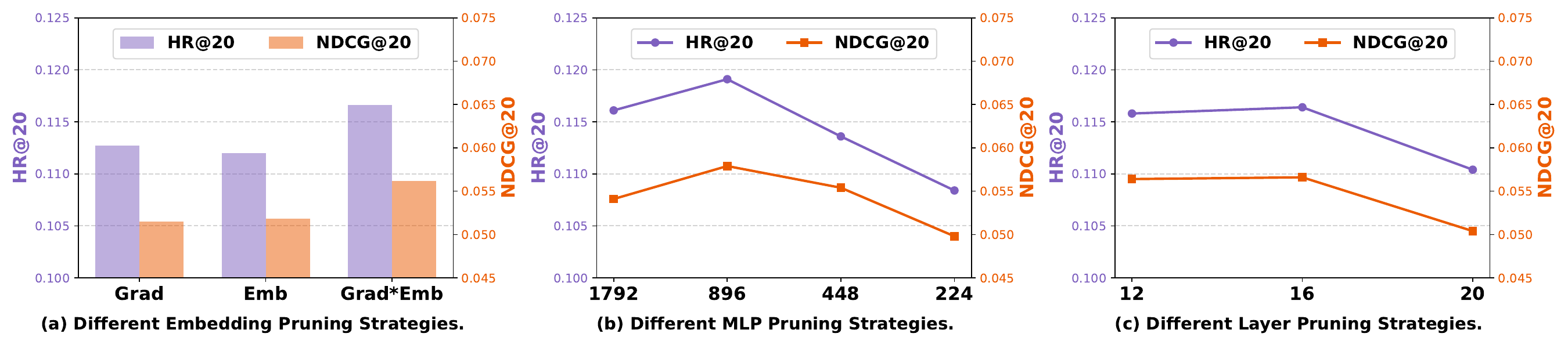} % 图片路径
  \vspace{-7pt}
  \caption{This figure presents the performance results of our analysis experiments on different modules. In each plot, the left vertical axis corresponds to the HR@20 metric, while the right vertical axis represents the NDCG@20 metric. Figure (a) compares the performance of various pruning strategies applied to the embedding module in Stage I. Figure (b) shows the performance impact of reducing the MLP intermediate size to different dimensions in Stage II. Figure (c) illustrates the performance comparison after pruning varying numbers of layers in Stage III.}                % 图片标题
  
  \label{fig:rq2}                                        % 标签（必须放在caption后）
\end{figure*}
\subsection{Main Results}

The main results of our work are presented in Table~\ref{tab:main_results}. From the table, we can draw the following observations:
\begin{itemize}[leftmargin=*]
    \item Compared to baseline models based on traditional recommender sysmtems, both DLLM2Rec and our method demonstrate significant performance advantages. This highlights that leveraging the knowledge of LLMs can greatly enhance recommendation capabilities.
    \item It is observed that SLMRec performs comparably or slightly worse than traditional methods, and its performance is significantly lower than our method, despite having a similar number of parameters. This unexpected result indirectly suggests that while layer parameters may be redundant, excessive pruning of layers can lead to a collapse in model performance.
    \item When comparing our method with DLLM2Rec, our approach shows a clear advantage. We attribute this to two main reasons: (1) We make full use of the original parameters of the LLM, preserving its intrinsic logic and knowledge to the greatest extent. (2) Thanks to the shared tokenizer, we are able to perform distillation in a softer logits space, further enhancing the effectiveness of knowledge transfer.
    \item Compared to the fully fine-tuned BIGRec, our method achieves the minimal performance loss across the three datasets, with an average drop of only 88\%, while utilizing only 5\% of the total parameters. This clearly demonstrates the significant improvement in recommendation effectiveness achieved by our approach.
    % 对比全量训练的模型，我们的方法在三个数据集上的性能损失最小平均只有XXX，而整体参数量只有XX\%，充分说明了我们方法对推荐效益的显著提升
\end{itemize}
% 1. 

% 2. It is observed that SLMRec performs comparably or slightly worse than traditional methods, and its performance is significantly lower than our method, despite having a similar number of parameters. This unexpected result indirectly suggests that while layer parameters may be redundant, excessive pruning of layers can lead to a collapse in model performance.

% 3. When comparing our method with DLLM2Rec, our approach shows a clear advantage. We attribute this to two main reasons: (1) We make full use of the original parameters of the LLM, preserving its intrinsic logic and knowledge to the greatest extent. (2) Thanks to the shared tokenizer, we are able to perform distillation in a softer logits space, further enhancing the effectiveness of knowledge transfer.
\begin{table}[t]
  \centering
  \caption{This table demonstrates the impact of different pruning strategies for the self-attention module on model performance. Here, ``random'' refers to random pruning, (1) denotes the method ``w/o layer-wise importance'', and (2) represents the approach ``w/o $\alpha$''.}
  % 这个表展示了使用不同的self-attention模块的剪枝策略对模型性能的影响其中 \textbf{random}表示随机裁剪,(1) 表示 w/o Layer wise Importance的方法，（2） 表示 w/o $\alpha$的方法.
  \label{tab:head_pruning_methods}
  % \vspace{-4pt}
  % \sisetup{table-format=2.2\%} % 设置百分比格式
  \begin{tabular}{lcc}
    \toprule
    \textbf{Pruning Strategy} & {\textbf{HR@20}} & {\textbf{NDCG@20}} \\
    \midrule
    random       & 0.1034 & 0.0435 \\
        w/o (1) + (2) & 0.1029 & 0.0424 \\
    w/o (1)& 0.1111 & 0.0530 \\
    % \cdashline{1-3}
    \rowcolor[HTML]{EFEFEF} PruneRec (ours)     & \textbf{0.1166} & \textbf{0.0562} \\
    \bottomrule
  \end{tabular}
% \vspace{em}
\end{table}

\subsection{Analysis}
% 为了进一步说明我们方法的每一步个模块的有效性，我们在这一章节中对每个模块分别进行了消融实验
To further demonstrate the effectiveness of each module in our method, we conducted an in-depth analysis of our method, guided by the following four research questions (RQs):
% \begin{itemize}[leftmargin=*]
%     \item \textbf{RQ1}. How do the model's parameter count and performance evolve across different stages of our method?
%     \item \textbf{RQ2.} What is the impact of our width pruning strategy of different modules during Stage I?
%     \item \textbf{RQ3.} How does varying the extent of parameter pruning in Stage II affect overall performance?
%     \item \textbf{RQ4.} What is the relationship between the number of pruned layers and the resulting model performance?
% \end{itemize}
\textbf{RQ1}: How do the model's parameter count and performance evolve across different stages of our method?
\textbf{RQ2}: What is the impact of our width pruning strategy of different modules during Stage I?
\textbf{RQ3}: How does varying the extent of parameter pruning in Stage II affect overall performance?
\textbf{RQ4}: What is the relationship between the number of pruned layers and the resulting model performance?
% \end{itemize}

\subsubsection{Performance on different stage (RQ1)}

To demonstrate the effectiveness of our model across the three stages, we present the performance of our three-stage pipeline on three datasets in Figure~\ref{fig:rq1}. From this figure, we can draw the following conclusions: (1) Pruning the model from different perspectives allows us to steadily maintain the model's performance level while continuously reducing its parameters.
(2) Pruning the Attn layers results in an average performance drop of 8\%, but reduces the parameters by about 83\%. Meanwhile, further pruning the MLP layers to 33\% of their original size allows the model to retain its performance level on average. This validates our preliminary experiments and indicates that these two modules contain significant information redundancy.

\subsubsection{In-depth analysis of each pruning strategy in Stage I (RQ2)}
 
\paragraph{Self-Attnention Module}
First, To show the effectiveness of our pruning method in the self-attention layer, we designed three sets of ablation experiments while ensuring that the pruning ratio remained consistent across all experiments:
(1) \textbf{w/o Layer-wise Importance}: Instead of computing importance scores for each layer individually, only a global importance score is calculated, and the same positions of attention heads are pruned across all layers.
(2)\textbf{w/o $\alpha$}: The value of $\alpha$ is fixed at 0, meaning that each layer only considers its own importance without accounting for cross-layer influences.
(3) \textbf{Random}: Randomly prune attention heads.

The experimental results are shown in Table~\ref{tab:head_pruning_methods}. First, it is evident that if  $\alpha$ is not emphasized, pruning attention heads leads to a significant drop in performance. Furthermore, if layer-wise importance estimation is removed and replaced with global estimation, the performance degrades to a level similar to random pruning. This demonstrates the effectiveness of our proposed pruning method.
In summary, the advantages of our approach can be highlighted in two key aspects: (1) We introduce the cross-layer influence factor $\alpha$, which establishes interde pendencies between shallow and deep attention heads. This allows pruning decisions in deeper layers to be guided by structural changes in shallower layers. (2) Our method retains the most task-relevant combination of attention heads in each layer, ensuring optimal performance while reducing redundancy.
\paragraph{Embedding Module} 
To tackle the issue of parameter redundancy in the embedding layer, we design two ablation experiments: (1) \textbf{Gradient Only}: This strategy quantifies the importance of weights based on the absolute value of the gradient during backpropagation.
(2) \textbf{Embedding only}: This approach measures the importance of weights using the absolute values of the embedding parameters.

Our results, presented in Figure~\ref{fig:rq2}, demonstrate that both pruning based solely on gradients and pruning based solely on embedding weights yield significantly inferior outcomes compared to our proposed method. We attribute our advantages to the following two points: (1) The gradient only strategy tends to prioritize weights that are highly sensitive during training but may overlook those that contribute to stable and generalizable feature representations. (2) The embedding only strategy focuses on retaining weights with large magnitudes, potentially neglecting smaller yet crucial parameters that play an important role in fine-grained distinctions.

\subsubsection{In-depth analysis of the impact of pruning strategy for MLP (RQ3).}

To further analyze and address the issue of parameter redundancy in the MLP layer during Stage II, we investigate the impact of pruning different proportions of parameters on model performance. As shown in Figure~\ref{fig:rq2}, we compare the model's performance when pruning the parameter count to 4×$d_{model}$ (1792), 2×$d_{model}$ (896), 1×$d_{model}$ (448), and 0.5×$d_{model}$ (224). The results indicate that the model achieves optimal performance when approximately 2×$d_{model}$ parameters are retained. Interestingly, increasing the dimensionality beyond this point does not necessarily lead to better performance, likely due to the negative impact of redundant parameters on the model’s effectiveness. On the other hand, reducing the dimensionality too aggressively causes a sharp decline in performance. These two observations indirectly validate our assumption: a significant portion of the model's parameters are redundant, and their presence can severely hinder practical deployment. However, excessively pruning these parameters can also harm the model's performance. Therefore, selecting a moderate pruning ratio is crucial for achieving an optimal balance.

\subsubsection{In-depth analysis of the impact of pruning strategy for layer (RQ4).}

To understand the impact of pruning different numbers of layers, and to further explain why SLMRec experiences a significant performance drop under the same parameter count. Specifically, we pruned the model by removing 12, 16, and 20 layers, respectively, and the results are shown in Figure~\ref{fig:rq2}. The experimental results demonstrate that the model's performance declines sharply once pruning exceeds a certain number of layers. This indicates that while deep compression is effective in reducing model size and accelerating inference, excessive compression can impair the model’s ability to handle complex recommendation tasks. Additionally, the results reveal that within a certain depth range, pruning can effectively remove redundant parameters while maintaining performance and significantly improving computational efficiency. These findings also validate our assumption: coarse-grained pruning of entire layers tends to excessively eliminate important parameters within those layers, leading to a notable degradation in model performance.

% \begin{table}[H]
%   \centering
%   \caption{Embedding Pruning Strategies Comparison}
%   \label{tab:pruning_methods}
%   % \sisetup{table-format=2.2\%} % 设置百分比格式
%   \begin{tabular}{lcc}
%     \toprule
%     \textbf{Pruning Strategy} & {\textbf{HR@20}} & {\textbf{NDCG@20}} \\
%     \midrule
%     Gradient only       & 0.1127 & 0.0515 \\
%     Embedding only  & 0.1120 & 0.0518 \\
%     Gradient*embedding & \textbf{0.1166} & \textbf{0.0562} \\
%     \bottomrule
%   \end{tabular}
% \end{table}

% \begin{table}[H]
%   \centering
%   \caption{MLP Pruning Strategies Comparison}
%   \label{tab:mlp_pruning_methods}
%   % \sisetup{table-format=2.2\%} % 设置百分比格式
%   \begin{tabular}{lcc}
%     \toprule
%     \textbf{Pruning Strategy} & {\textbf{HR@20}} & {\textbf{NDCG@20}} \\
%     \midrule
%     % random       & 0.1145 & 0.0531 \\
%     % Wanda  & 0.0013 & 0.0004 \\
%     Top 1792      & 0.1161 & 0.0541 \\
%     Top 896      & 0.1191    & 0.0579 \\
%     Top 448      & 0.1136 & 0.0554 \\
%     Top 224 & 0.1084 & 0.0498 \\
%     \bottomrule
%   \end{tabular}
% \end{table}

% \begin{table}[H] %替换为4_layer_difference.png
%   \centering
%   \caption{Layer Pruning Strategies Comparison}
%   \label{tab:layer_pruning_methods}
%   % \sisetup{table-format=2.2\%} % 设置百分比格式
%   \begin{tabular}{lcc}
%     \toprule
%     \textbf{Layer numbers pruned} & {\textbf{HR@20}} & {\textbf{NDCG@20}} \\
%     \midrule
%     12       & 0.1158 & 0.0564 \\
%     16  & 0.1164 & 0.0566 \\
%     20      & 0.1104 & 0.0504 \\

%     \bottomrule
%   \end{tabular}
% \end{table}
% \end{document}
\section{Related Work}

In this section, we discuss the related works on model pruning and LLM-based recommendation.

\vspace{+5pt }
% \subsection{Model Pruning}
\noindent$\bullet$ \textbf{Model Pruning}.
In recent years, the demand for parameter compression and inference acceleration in large language models (LLMs) has become increasingly urgent~\cite{zhao2023survey, zhu2024survey}. Pruning techniques have emerged as important tools for removing redundant parameters from such large models. In the domain of model pruning, the main approaches include unstructured pruning, dynamic parameter pruning, and structured pruning.
Unstructured pruning removes low-importance weights by analyzing their individual importance, which also causes unstructured sparsity and hard deployment~\cite{sunsimple}. Although these methods can theoretically achieve high sparsity and significantly reduce parameter counts, they often ignore the overall structure of LLMs. 
Dynamic parameter pruning selects an appropriate sub-model for each input based on its specific characteristics.
Structured pruning, in contrast, removes redundant network structures such as entire neurons, channels, or layers~\cite{wang2019structured, ma2023llm, an2024fluctuation}. For example, works like ShortGPT~\cite{men2024shortgpt} perform compression by pruning redundant layers. SparseGPT~\cite{frantar2023sparsegpt} works by reducing the pruning problem to a set of extremely large-scale instances of sparse regression. LLM-Pruner~\cite{ma2023llm} leverages gradient information to remove redundant structures, LaCO~\cite{yang2024laco} compresses models through layer merging. However, all of these methods are designed for general-purpose tasks and are not specifically tailored to recommendation systems. Our approach introduces a structured pruning method that is explicitly designed to adapt LLMs for recommendation tasks.

\vspace{+5pt}
\noindent$\bullet$ \textbf{LLM for Recommendation}.

Impressed by the powerful capabilities of LLMs, many researchers have begun exploring using LLMs in recommendation tasks and achieved great recommendation performance~\cite{recsurvey1,recsurvey2,recsurvey3,recsurvey4}, which can be broadly categorized into two main types. 

The first approach uses LLMs to assist recommender systems, such as extracting semantic embeddings from user behavior sequences or item descriptions~\cite{yuan2023go, wei2024llmrec}, or leveraging LLMs' open-world knowledge to enhance the textual representation of items and user behaviors~\cite{xi2024towards}.

The second approach uses LLMs directly as recommenders to explore their recommendation potential.

Some early studies tried using the in-context learning ability of LLMs to do recommendation tasks directly~\cite{gao2023chat, hou2024large, dai2023uncovering}. However, as LLMs are not specifically designed for recommendation tasks, this approach has shown limited effectiveness~\cite{bao2023tallrec}. 

Therefore, many studies focus on instruction tuning for LLMs in recommendation tasks to unleash their recommendation potential and achieve strong performance in recommendation~\cite{bao2023bi, zhang2025collm, zhang2023recommendation}, which is the LLM for recommendation direction this paper emphasizes.

However, directly deploying LLMs for recommendation is costly. Two methods have been raised to address this: one reduces redundant layers via model pruning and knowledge distillation~\cite{xu2024slmrec} but overlooks redundancy in key components like self-attention and MLP layers; the other transfers LLMs' capabilities to lightweight traditional recommendation models through knowledge distillation~\cite{cui2024distillation}, but the capacity gap between teacher and student models makes it hard for traditional recommender models to fully capture the LLM's recommendation-related information.

In contrast to prior methods, this paper introduces a recommendation-specific iterative ``prune-and-restore'' pipeline that minimizes model parameters while maintaining recommendation performance by pruning redundancy in self-attention and MLP layers, balancing parameter efficiency and recommendation effectiveness.
% }

\section{Conclusion}

% In detail, in th
% 在这篇文章中我们提出了PruneRec，我们首先指出LLM-based recommender system 受限于LLM本身并不是为了推荐任务设计的，其内部存在大量的参数服务于诸如math，code等任务，which对于推荐任务来说增加了参数量和开销，但是很难带来实质性的收益。To over come this problem, 我们conduct 了观察实验发现在现有的LLM在结构上对于推荐任务存在大量的冗余参数。为此我们提出了PruneRec，一个pipeline specifically designed for LLM- based recommender system. 通过三阶段的裁剪和修复在裁剪了95\%参数的同时，平均保留了接近90\%的性能，极大的降低了LLM-based Recommender Systm的部署开销

% In this paper, we begin by highlighting that LLMs are not inherently designed for recommendation tasks, and as a result, they contain a significant number of parameters tailored for unrelated tasks such as mathematical reasoning or code generation. These parameters increase both the computational cost and memory overhead without providing meaningful benefits for recommendation performance. To tackle this issue, we conducted exploratory experiments and observed substantial structural redundancy in LLMs when applied to recommendation tasks. Motivated by these findings, we propose PruneRec, a pipeline specifically designed for LLM-based recommender systems. Through a three-stage pruning and refinement process, PruneRec achieves an impressive reduction of 95\% in parameter count while retaining nearly 90\% of the original performance on average. This dramatic reduction significantly lowers the deployment costs of LLM-based recommender systems, making them more practical and efficient for real-world applications.

In this paper, we first emphasize that LLMs are not originally designed for recommendation tasks. Consequently, they include a significant number of parameters optimized for unrelated functions, such as mathematical reasoning or code generation. These parameters increase computational costs and memory usage without offering meaningful improvements in recommendation performance. To address this challenge, we conducted exploratory experiments and identified substantial structural redundancy in LLMs when applied to recommendation tasks. Inspired by these insights, we propose \textbf{PruneRec}, a pipeline specifically tailored for LLM-based recommender systems. Through a three-stage pruning and refinement process, PruneRec achieves an impressive 95\% reduction in parameter count while retaining nearly 90\% of the original performance on average. This dramatic reduction significantly decreases the deployment costs of LLM-based recommender systems, enhancing their practicality and efficiency for real-world applications.

% \section{Problem statement}

% In this document we discuss how to write an ACM article.

% \section{Methods}

% This document provides \LaTeX\ templates for the article. We
% demonstrate different versions of ACM styles and show various options
% and commands.  We add extensive documentation for these commands and
% show examples of their use.

% % \section{Experiments}
% % \subsection{Experimental Setup}
% % \paragraph{Models and Datasets} We primarily evaluate out method using Qwen2-0.5B\cite{}.
% % \input{samples/4Experiments}

% \section{Results}

% We hope the resulting templates and documentation will help the
% readers to write submissions for ACM journals and proceedings.

% \section{Significance}

% This document is important for anybody wanting to comply with the
% requirements of ACM publishing.

\bibliographystyle{ACM-Reference-Format}
\bibliography{main}

\end{document}